\documentclass[pre,twocolumn,showpacs,floatfix,openany]{revtex4-1} 

\usepackage{graphicx}
\usepackage{dcolumn}
\usepackage{amsmath}
\usepackage{epsfig}
\usepackage{amssymb}
\usepackage{wrapfig}
\usepackage{sidecap}
\usepackage{xfrac}
\usepackage{footnote}
\usepackage{caption}
\usepackage{natbib}
\usepackage{float}
\usepackage{color}
\usepackage{lipsum}
\usepackage{footnote}
\usepackage{tabularx}

\newcommand{\beq}{\begin{equation}}
\newcommand{\eeq}{\end{equation}}
\newcommand{\bea}{\begin{eqnarray}}
\newcommand{\eea}{\end{eqnarray}}

\newcolumntype{C}[1]{>{\centering\arraybackslash}p{#1}}

\begin{document}

\title{Effect of surface morphology on kinetic compensation effect}

\author{Nayeli Zuniga-Hansen}
\affiliation{Department of Physics \& Astronomy, Louisiana State University,
  Baton Rouge, Louisiana 70808, USA}
\email{zunigahansen@lsu.edu} 

\author{Leonardo E. Silbert}
\affiliation{School of Math, Science, and Engineering, Central New Mexico
  Community College, Albuquerque, New Mexico 87106, USA}

\author{M. Mercedes Calbi}
\affiliation{Department of Physics, University of Denver, Denver, Colorado
  80208, USA}

\begin{abstract}

  As part of a systematic study on the kinetic compensation effect, we use
  kinetic Monte Carlo simulations to observe the effects of substrate topology
  on the transient variations in the Arrhenius parameters - effective
  activation energy $E_{a}$, and preexponential factor $\nu$ - during thermal
    desorption, with a particular focus on differences between ordered and
    disordered surfaces at a fixed global coordination number. The rates of
    desorption depend on surface configuration due to the inherent differences
    in the local environments of adsorbing sites in the two cases. While the
    compensation effect persists for the disordered substrate, the change in
    topology introduces an element that produces variations in $\nu$ that are
    independent of variations in $E_{a}$, which implies that the parameters
    cannot be fully characterized as functions of each other. We expect our
    results to provide a deeper insight into the microscopic events that
    originate compensation effects in our system of study but also in other
    fields where these effects have been reported.

\end{abstract}

\maketitle



\section{INTRODUCTION}

The kinetic compensation effect (KCE), observed in many different areas of the
physical, chemical, and biological sciences, is the systematic variation in the
\textit{apparent magnitudes} of the Arrhenius parameters, the energy of
activation $E_a$, and the preexponential factor $\nu$, as a response to a
change in an experimental parameter in a set of closely related activated
processes. The extracted values of $E_{a,j}$ and $\ln{\nu_j}$ from the
$j^{th}$ Arrhenius plot in the series are often observed to satisfy a linear
relation of the form, 
\beq
\ln{\nu_j} = \beta E_{a,j} +\ln{k_o}
\label{eq:IKR}
\eeq 
for constant $\beta$ and $k_o$
\cite{LiuGuo:01,Perez:16,Freed:11,Lvov:13,Pan:15,Zuniga:18,Barrie:12a,
  Barrie:12b,Yelon:12, Kreuzer:88}.

\begin{figure}[h!]
(a)
    \includegraphics[clip,width=1.5in]{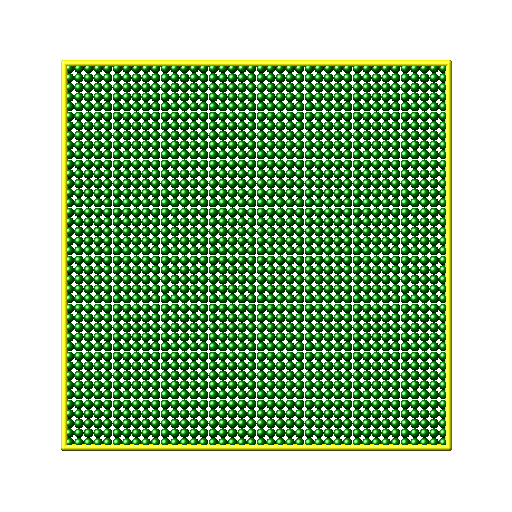}
(b)
    \includegraphics[clip,width=1.45in]{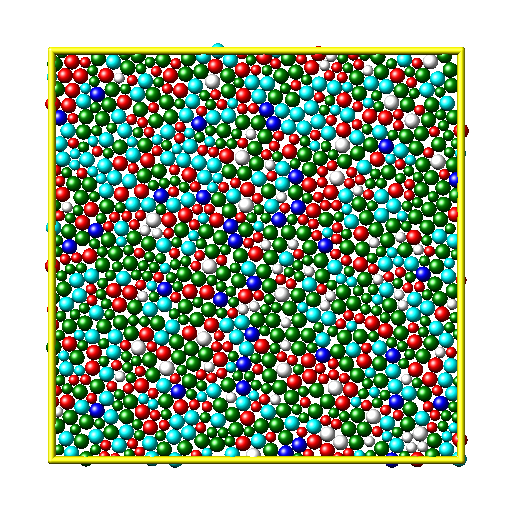}
    \caption{The two bidimensional surfaces studied in this work, each
      possessing a global average coordination number equal to four. (a)
      Perfect square lattice containing 1600, monodisperse sites. (b)
      Disordered configuration composed of a 50:50 mixture of 1000 sites of
      size ratio 1:4 (to suppress ordering during the configuration
      preparation protocol). Local site coordination is indicated by particle
      shading. Green equals a contact number of 4. The redder end of the spectrum
      represents lower number of nearest neighbors, whereas bluer shades are
      higher. Note: white/gray sites have zero contacts.}
\label{fig:config}
\end{figure}

The implication of the term compensation is that any effect of variations in
$E_a$ on the rate of the process, given by the Arrhenius equation 
$k= \nu \exp{\frac{E_a}{k_B T}}$, where $k_B$
is the Boltzmann constant and $T$ the temperature, are offset by variations
in the prefactor $\nu$, in the same direction and with the same, or almost
the same magnitude, such that $k$ remains relatively unchanged
\cite{Estrup:86,LiuGuo:01,Piguet:11,Miller:87,Tomkova:98,Yelon:12,Dunitz:95,
  Tomkova:98,Gottstein:98,Pan:15}.

The slope of Eq.~\ref{eq:IKR} sometimes yields a temperature, called the
\textit{compensation temperature} $T_c$, as $\beta = \frac{1}{k_B T_c}$, at
which a set of $\ln{k_j}$ vs. $\frac{1}{k_B T}$ plots are observed to cross
and the rates $k_j$ acquire the same value $k_j = k_o$, and are said to become
unaffected by external perturbations \cite{LiuGuo:01,Yelon:12,Pan:15,Lvov:13,Barrie:12a,Barrie:12b,Freed:11}.
This convergence
is called the \textit{isokinetic relation} (IKR). The IKR is often mentioned
interchangeably with the KCE \cite{Pan:15,LiuGuo:01}, perhaps because the
occurrence of an IKR is often attributed to the parameters compensating each
other at $T_c$ \cite{Douglas:09,LiuGuo:01}, however the authors in
\cite{LiuGuo:01} note that this mutual correspondence is only exact if the
linear correlation coefficient between data points in Eq.~\ref{eq:IKR} is $1$.

The KCE and IKR are generally identified as features of weak molecular
interactions \cite{Dunitz:95, Piguet:11, Ford:05} and, along with the closely
related entropy-enthalpy compensation, have been reported in a wide range of
phenomena in the physical, chemical, and biological sciences
\cite{LiuGuo:01,Freed:11}, yet continue to be the subject of debate as to
their existence and validity. Some authors claim that the KCE is a consequence
of the exponential nature of the Arrhenius rate equation
\cite{Koga:91,Cornish:02,Barrie:12a}.  The observation of a strong linear
correlation is commonly attributed to the parameters being extracted from the
same temperature dependent data
\cite{LiuGuo:01,Barrie:12a,Cornish:02,Perez:16}, or deemed a consequence of
experimental and/or statistical errors \cite{Barrie:12a,Barrie:12b,Perez:16}.

In Ref.~\onlinecite{Zuniga:18} we studied the kinetic compensation effect
during the desorption of interacting and noninteracting adsorbates from an
energetically homogeneous crystalline, square lattice surface. That previous
study showed that the parameters exhibit a rather weak \textit{partial
  compensation effect} in all regimes of interaction strength, because
variations in $\nu$ are not strong enough to offset those in $E_a$. In
addition, the observed IKR was found to be due to the system transitioning to
a non-interacting regime, not because of a mutual offsetting between the
Arrhenius parameters.

In the present work, we compare the results from the square lattice with those
for the thermal desorption from a two dimensional disordered or amorphous
surface. This is part of a systematic study where we implement a kinetic Monte
Carlo scheme to numerically calculate the transient behavior of $E_a$
throughout the thermal desorption process of interacting and noninteracting
adsorbates when different `experimental' parameters are altered.  The
calculated data allow us to also obtain the transient variations in the
prefactor, in order to quantify the level of compensation between $E_a$ and
$\nu$.  This approach differs from previous ones
\cite{Estrup:86,Piguet:11,Tomkova:98,Dunitz:95, Tomkova:98,Starikov:07}, in
the sense that it does not part from the assumption that $E_a$ and $\ln{\nu}$
must satisfy the strong linear correlation in Eq.~\ref{eq:IKR}, nor
preconceived functional forms based on it.  In each set of results the net
attractive interaction strength is the same, while the parameter that is being
altered is the surface topology.  These results also allow for an overview of
the kinetics of desorption from amorphous surfaces, which constitutes an
important problem, since disorder is present in many realistic systems, and
exact functional forms can be difficult to obtain \cite{Talbot:07,Talbot:08}.

\section{Methodology} 

We use the kinetic Monte Carlo algorithm \cite{Voter:book} to simulate the
thermal desorption of interacting and noninteracting, quasispherical
adsorbates from the bidimensional surface configurations shown in
Fig.~\ref{fig:config}. This study focuses on the comparison and contrast
between the two particular surfaces: a two-dimensional square lattice and a
two-dimensional disordered surface. The main distinction between them is the
distribution of site coordinations. For the square lattice
(Fig.~\ref{fig:config}(a)) each site has exactly four neighbors. Therefore,
the global average coordination number for the `ordered' square lattice is
also precisely four, $z_{O} = 4$. Whereas, for the disordered configuration
(Fig.~\ref{fig:config}(b)), the local site coordination number is not constant
such that there are varying numbers of nearest neighbors from site to site,
spanning 0 to 6. This variation in site coordination is indicated by the
particle shading in Fig.~\ref{fig:config}.  However, the disordered surface
has been prepared such that it's average coordination number matches that of
the square lattice, i.e. $z_{D} = 4$. The disordered surface is fully
representative of a realistic disordered system, such as a connected layer of
sand grains or an amorphous, glassy substrate \cite{coniglio1}.

For this study, the lattices are energetically homogeneous, with binding
energy $E_{i} = E_{b}= 100$, in units where $k_{B}=1$. The index indicates the
$i^{th}$ site on the surface. Lateral adsorbate-adsorbate interactions are
added as a parameter $\epsilon$, which is calculated as a percentage of
$E_b$. The interaction strengths employed in this study are $0\%$, $10\%$,
$50\%$ and $90\%$ of the binding energy.

To track the desorption process, the kinetic Monte Carlo algorithm follows a
series of steps. First, initial conditions are specified. This includes
binding and interaction energies ($E_b$ and $\epsilon$, respectively), initial
temperature $T_0$ (which is modified depending on $\epsilon$), step size for
temperature increase $\gamma$, as $T = T_0 + \gamma t$, and initial coverage,
here set at monolayer ($100\%$) for all cases.  During the second step the
algorithm calculates the number of occupied nearest neighbors per site, as
well as site energies in order to determine transition probabilities. These
are calculated as $W_i = e^{\beta E_i}$, where $E_{i}$ is the energy barrier
of the $i^{th}$ adsorption site, given by
\beq 
E_{i} = E_{b} + \sum_{m = 1}^{z_{i}} n_{im} \epsilon,
\nonumber
\eeq 
where each site $i$ picks up an energy contribution from its $z_{i}$ nearest
occupied neighbors. Thus, $n_{im} = 1$ when a neighbor site is
occupied, and is zero if empty.

Next, a random number $x_1$ between $0$ and $1$ is generated, to select an
allowable transition (desorption or diffusion to a neighboring available
site).  The selected change of state is that with the largest probability,
which satisfies the following inequality
\begin{equation}
\frac{1}{W} \sum_{j=1}^{k-1} W_{ij} <x_1 < \frac{1}{W} \sum_{j=1}^{k} W_{ij},
\nonumber
\end{equation}
where $W_{ij}=e^{\beta(E_i - E_j)}$ is the transition probability from state
$i$ to state $j$.  The sum $W$ of all probabilities is over all $k$ allowed
transitions per site $i$.  Lower probability transitions can still take place,
to allow the system to evolve freely, and avoid forcing it to follow a
particular path.  After every transition, the time variable $t$ increases by a
fractional amount determined by a second random number. Temperature increases
according to the step size $\gamma$, here set to $1$ degree per unit of time.
The average site occupancy and energy are recorded at every iteration, and the
process is repeated until all the particles have desorbed. The results are
obtained as an (ensemble) average over many independent runs. 

The data analysis is done with the Polanyi-Wigner equation for desorption,
\beq
-\dot{\theta} = \theta^n \nu \exp{\frac{-E_a}{k_B T}}
\label{eq:PW}
\eeq 
where $\theta$ is the fractional coverage and $\dot{\theta}$ its time
derivative, $n$ is the order of the process, which is set to $n =1$ for
reversible thermal desorption. It is worth pointing out that Eq.~\ref{eq:PW} can
only be fit exactly in the noninteracting regime, where $E_a$ and $\nu$ remain
constant, or if their functional forms are known. In the present work the
numerical data for $E_a$, along with $\dot{\theta}$ are used to extract the
nonconstant preexponential factor.

\section{RESULTS}

\subsection{Rates of desorption and Arrhenius plots}

The first series of results consists of a comparison between desorption rate
curves, and corresponding Arrhenius plots from both surfaces. In each set the
interaction strength is the same, and the parameter that is being altered is
the surface morphology.  The values of the interaction energy $\epsilon$ for
the data sets presented in this series are $0\%$, $10\%$, $50\%$ and $90\%$ of
the binding energy, $E_b = 100$. (In the proceeding figures we plot the
magnitude of $E_{a}$, as the activation energy itself is negative.)
\begin{figure}
    \includegraphics[width=3.5in]{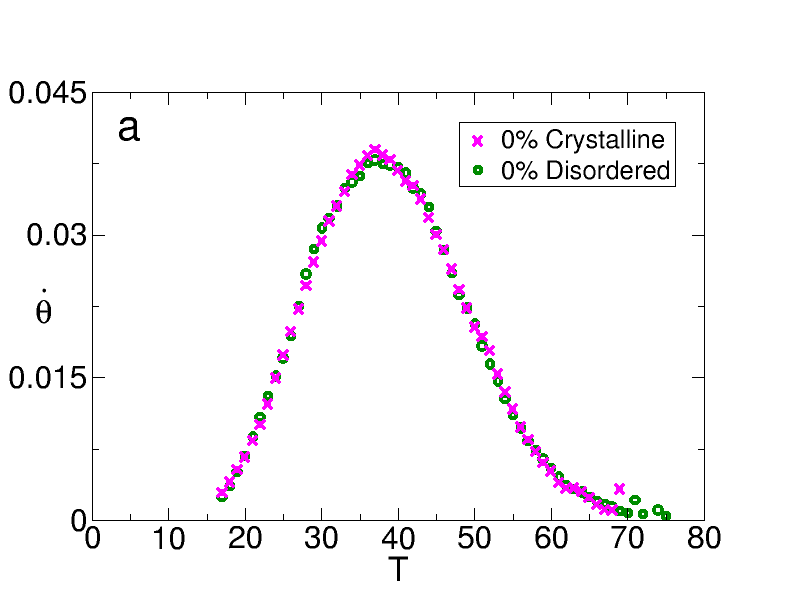}
    \includegraphics[width=3.5in]{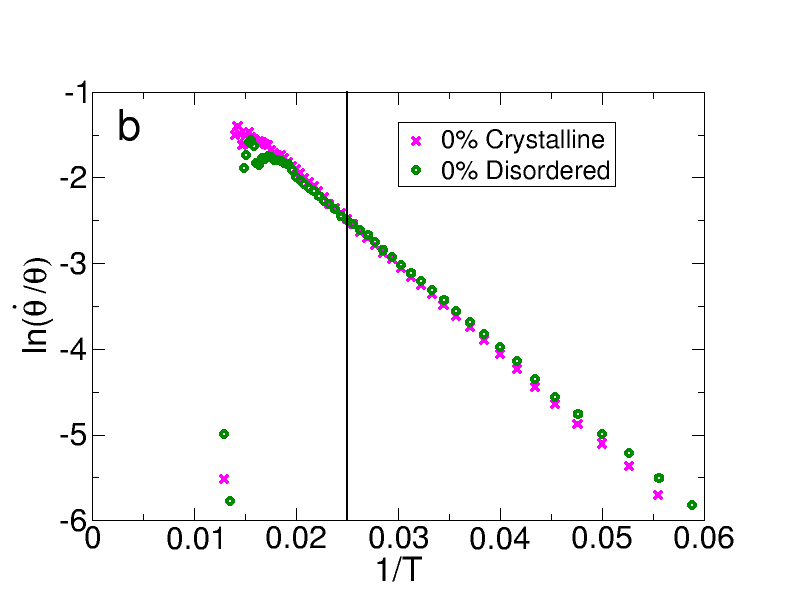}
    \caption {Thermal desorption rates from both the crystalline (ordered)
      and disordered surfaces for noninteracting species. (a)
      Desorption rate $\dot{\theta}$ and (b) corresponding Arrhenius plot. 
      The solid black line in (b) indicates the crossing temperature of the Arrhenius
      plots.}
\label{fig:Noint}
\end{figure}

In the non-interacting regime (Fig.~\ref{fig:Noint}) the desorption rate
curves from both surfaces in Fig.~\ref{fig:Noint}(a) exhibit very little
differences, since the curves almost overlap completely for the entire
process.  The corresponding Arrhenius plots in Fig.~\ref{fig:Noint}(b) also
exhibit significant overlap, nevertheless, a small gap at low temperatures can
be seen upon closer inspection.  The plots come closer together as $T$
increases, and cross at approximately $T = 40$ (or $\frac{1}{T} = 0.025$). 
This could be an isokinetic
relation in the traditional sense, however, it cannot be associated with a
compensation effect, since the activation energy is constant and cannot
influence the preexponential factor. Nevertheless, the rates, $k \equiv
\frac{\dot{\theta}}{\theta}$, for both surfaces acquire close values at $T =
40$, where their ratio (ordered / disordered) is $k_{O}/k_{D} \approx 0.985$.

A thermodynamic interpretation of the Arrhenius parameters states that the
prefactor has an entropy component, $ e^{\frac{\Delta S}{k_B}}$, where $\Delta
S$ is the change in the entropy, and a frequency component, $\kappa$
\cite{Sharp:01,Gottstein:98,Piguet:11}, of attempted events, in this case of
desorption events.  Sometimes an additional factor $\rho$ is considered, which
corresponds to geometric constants of the system in question
\cite{Gottstein:98}.

In the non interacting regime the observed differences can be attributed to
the frequency of desorption events.  This is because the disordered surface
has sites with varying numbers of nearest neighbors, $ 0 \leq z_{i} \leq
6$. Sites with $z_{i} = 0$, the `rattlers', have no nearest neighbors,
and thus a particle initially located there is only allowed to desorb. This
is therefore the initial step in the desorption process for the disordered surface.
For sites with larger coordination number, connectivity to neighboring sites
allows particles to diffuse to a nearest available location. This tends to be
the preferred transition at lower temperatures, and causes particles to linger
on the surface slightly longer. In addition, sites with larger values of
$z_{i}$ are more accessible to particles that remain on the surface, making
their reoccupation easier, which also contributes to their coverage decreasing
at a slower pace.  There is also a decrease in configurational entropy arising
from the various coordination numbers.  Once particles desorb from rattler
sites there is no probabilty for reoccupation, since, in this study, we
exclude readsorption, mimicking the process whereby desorbed particles are
extracted from the chamber in an experiment. Additionally, sites with lower values of $z_i$ are
also less accessible once unoccupied than those with more nearest
neighbors. This limits the number of ways that the remaining particles can be
distributed among the available locations in the amorphous configuration.  These factors add
complexity to the desorption process, as they result in multiple desorption
rates. In this particular configuration they produce a slightly different
value of $\ln{k} \equiv \ln\left(\frac{\dot{\theta}}{\theta}\right)$ from that in the
crystal at low $T$, since the overall curve is the average of all those
contributions.
\begin{figure}
    \includegraphics[width=3.5in]{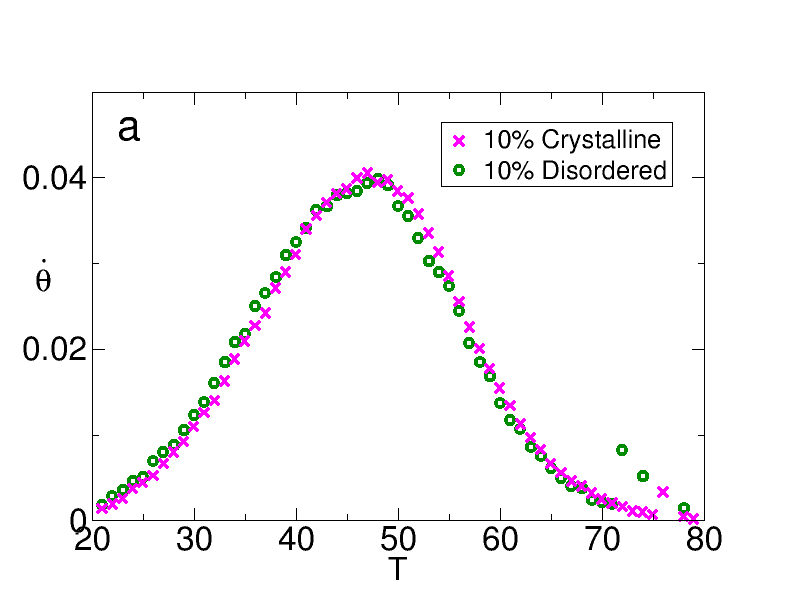}
    \includegraphics[width=3.5in]{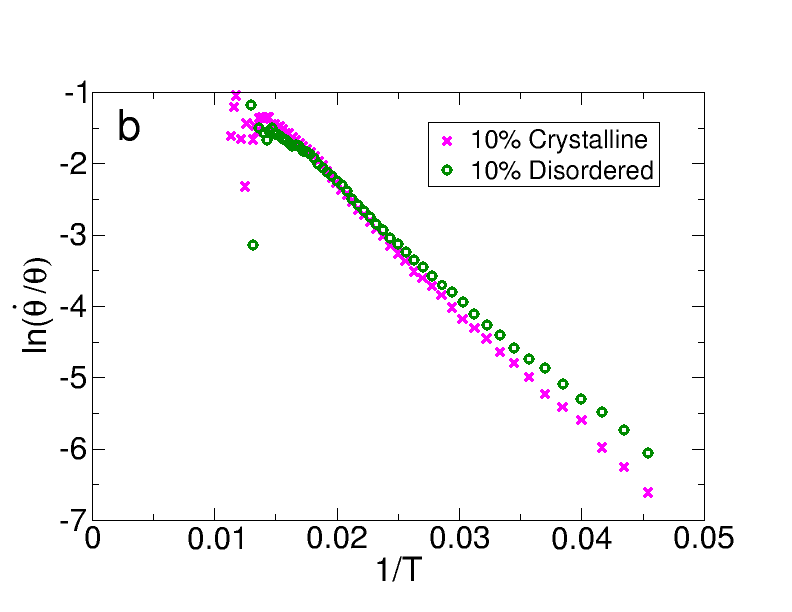}
    \caption{Thermal desorption rates from the crystal and the disordered
      surfaces (a), and corresponding Arrhenius plots (b), at $10\%$
      interaction strengh. The Arrhenius plots in this regime exhibit the
      behavior associated with an isokinetic relation (IKR). The crossing
      temperature is visually estimated at $T \approx 47$. The KCE is
      associated with weak molecular interactions, and the fairly linear
      Arrhenius plots allow the extraction of apparently constant values of
      $E_a$ and $\ln{\nu}$.}
\label{fig:10int}
\end{figure}

Even at only $10\%$ interaction strength, differences in the desorption
rates become more evident, as shown in Fig.~\ref{fig:10int}(a), but
nevertheless remain small. These differences should be expected, since lateral
interactions, combined with varying site values of $z_{i}$ for the disordered
surface, add energetic heterogeneity to the lattice, so that differences
between sites are enhanced.

The corresponding Arrhenius plots in Fig.~\ref{fig:10int}(b) also exhibit more
noticeable differences. The initial gap between them is larger, and they
eventually converge and exhibit an IKR. For this data set, the temperature of
greatest overlap between the ordered and disordered surfaces, shown in
Fig.~\ref{fig:10int}(b), occurs at $T \approx 47$, at which point the
Arrhenius parameters almost match exactly, where the fractional surface coverage
reaches $50\%$.  Even though the parameters are numerically close, this
contrasts with the results of Ref.~\cite{Zuniga:18}, where the overlap
occurred when the system transitioned to the non-interacting regime. Here, at
the crossing point, lateral interactions still have some effect, and it
appears that the reason for the convergence is that the effective average
coordination number $z_i$ occupancy per site becomes almost the same in both
surfaces, at $z \approx 1$ (as seen in the values of $E_a$). 
This presents a different scenario where an IKR is
observed, but one \textit{which also precludes the occurrence of complete
  compensation between $E_a$ and $\ln{\nu}$}.

The IKR observed here, as well as that in Ref.~\cite{Zuniga:18}, for this same
weak interaction regime ($\le 10\%$), which is accompanied by a strong linear
correlation that fits Eq.~\ref{eq:IKR}, shows why these phenomena are usually
ascribed to weak molecular interactions \cite{Sharp:01, Chodera:13, Ford:05,
  Piguet:11}.

\begin{figure}
    \includegraphics[width=3.5in]{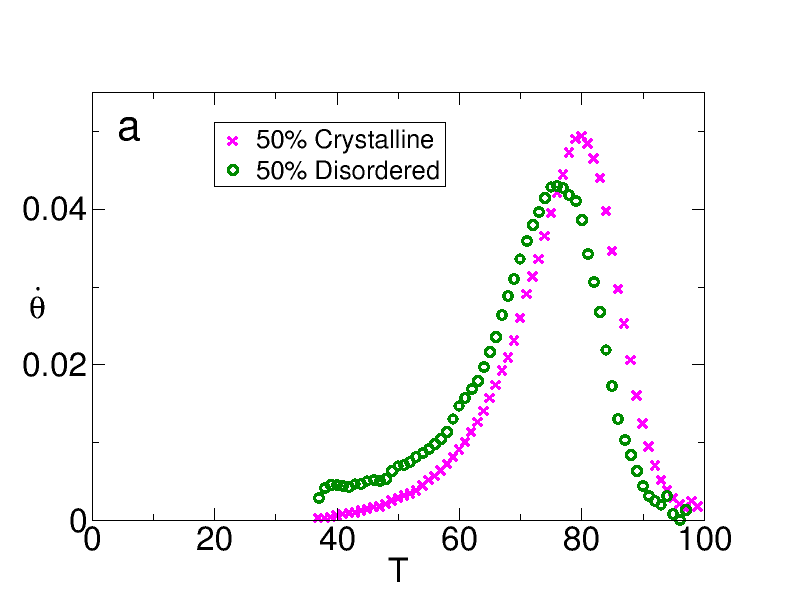}
    \includegraphics[width=3.5in]{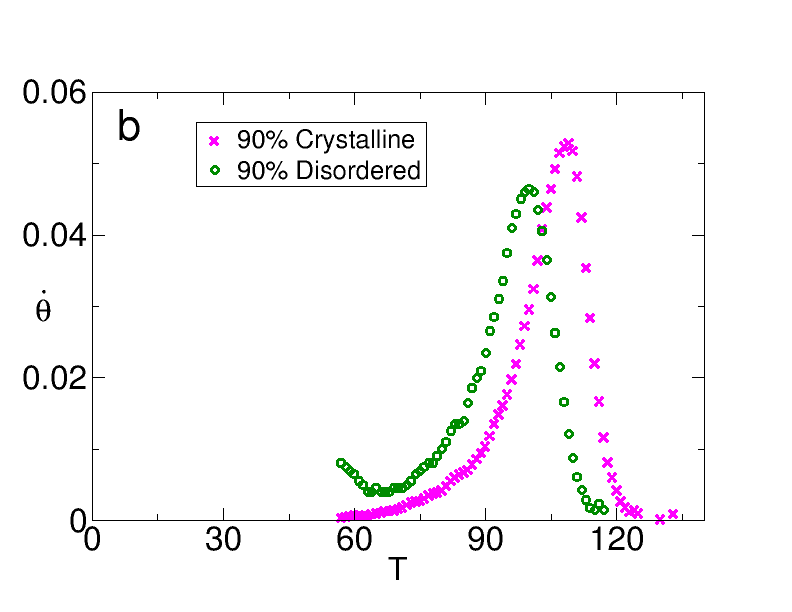}
\caption
{Comparison of thermal desorption rates from the crystal and the amorphous surface
 at (a) $50\%$ and (b) $90\%$ interaction strength. The
  desorption rate increases in the amorphous surface where multiple rates
  arise due to varying coordination numbers $z_{i}$ from site to site.}
\label{fig:5090int}
\end{figure}

\begin{figure}
    \includegraphics[width=3.5in]{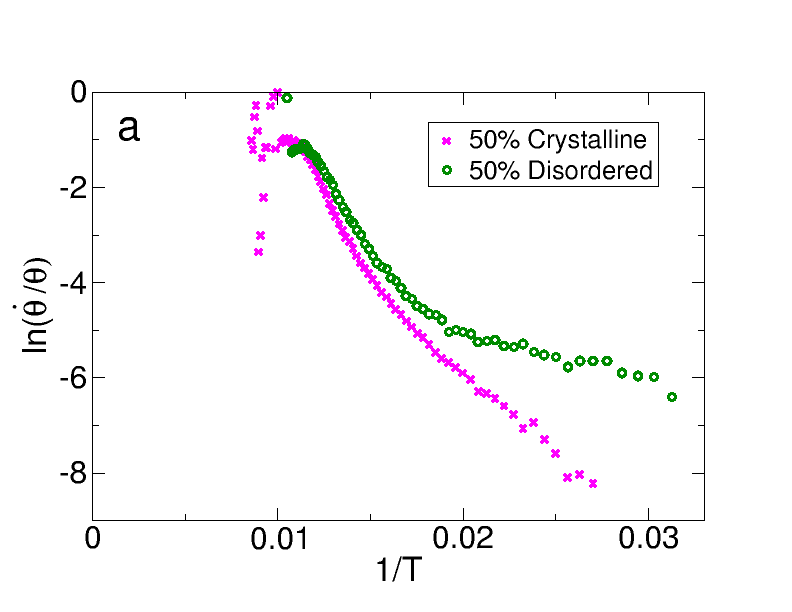}
    \includegraphics[width=3.5in]{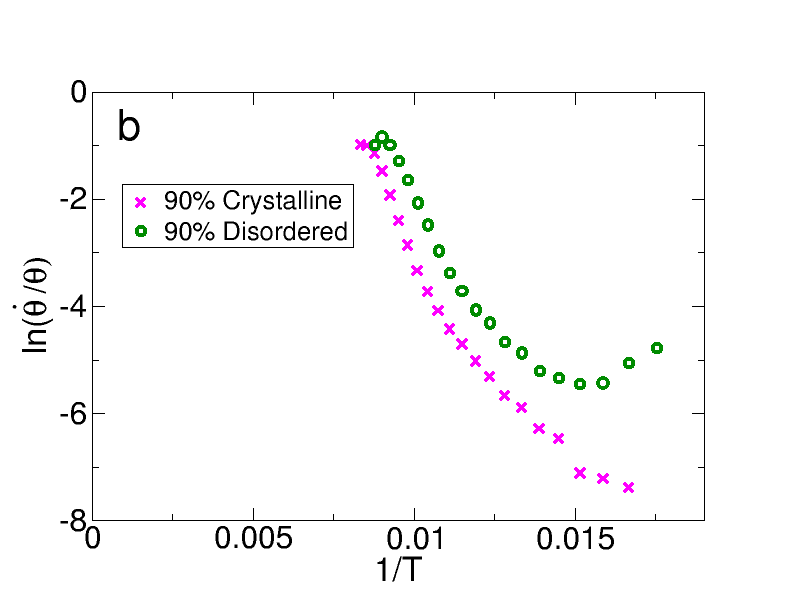}
\caption
{Comparison between Arrhenius plots from the crystal and the amorphous surface, 
 at (a) $50\%$ and (b) $90\%$ interaction strength. In the disordered configuration the 
curvature of the Arrhenius plots becomes more pronounced as interaction strength increases.
The plots appear to come together at high values of the temperature
where both surfaces are almost empty.}
\label{fig:5090aplots}
\end{figure}

In the $50\%$ and $90\%$ interaction strength regimes, 
the desorption rates in Figs.~\ref{fig:5090int}(a) and (b), respectively
exhibit even
more pronounced differences.  The rate of desorption is visibly faster in the
disordered surface and peaks at a lower temperature  There is also a visible
`tail' on the left end of the thermal desorption peak at $90\%$
interaction strength (see Fig.~\ref{fig:5090int}(b)) from the disordered
surface. This feature does not appear at $50\%$ interaction strength, but the
thermal desorption peak does not start at $0$ on the abscissa. The
corresponding Arrhenius plots in Figs.~\ref{fig:5090aplots}(a) and (b)
exhibit sub Arrhenius type behavior \cite{Silva:13}, i.e. a
concave curvature, signature of a variable energy of activation
\cite{Vyazokin:16}, but the curvature becomes significantly more pronounced for the
disordered configuration.

\subsection{Kinetics of desorption from an amorphous surface: an overview}

The shape of the Arrhenius plot is determined by the Arrhenius parameters
$E_a$ and $\ln{\nu}$, however, their transient variations, as well as some of
the features observed in the desorption rate peaks and Arrhenius plots in
the amorphous configuration, can be explained by looking at the individual
contributions from each group of sites, with specific $z_{i}$ values, to the
overall rates of desorption, as an initial overview to a more extensive study
on the kinetics of desorption from two dimensional amorphous lattices.

\begin{figure}
    \includegraphics[width=3.5in]{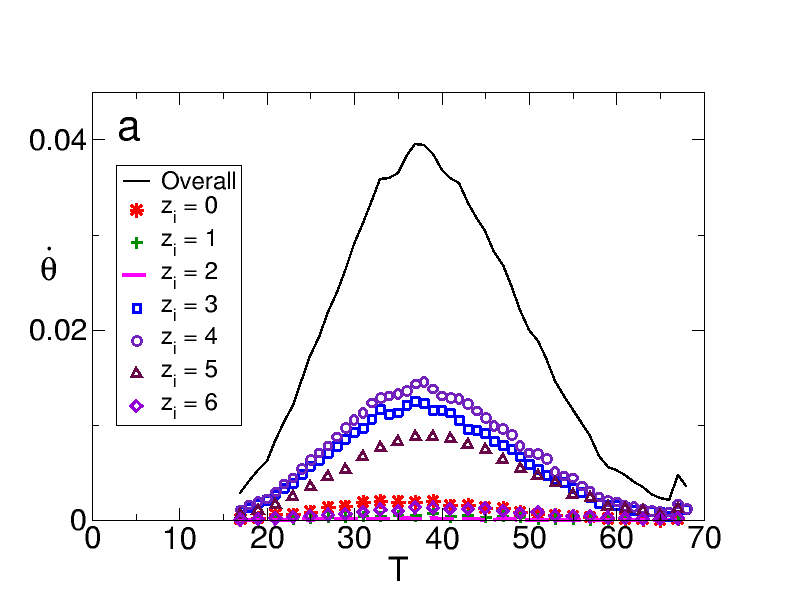}
  \includegraphics[width=3.5in]{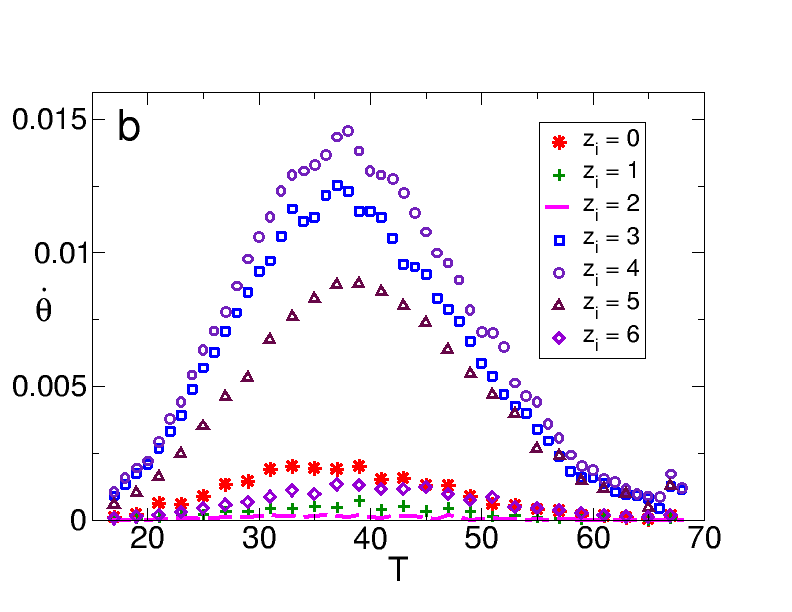}
  \caption{Desorption rates as function of temperature at $0\%$
    interaction strength. (a) Overall rate, and contributions for each
    group of sites, classified according to their coordination number
    $z_{i}$. (b) Magnified view of the site contributions, the peak from sites
    with $z_{i} = 0$ reaches a maximum at a slightly lower temperature than
    other sites, indicating faster desorption.}
\label{fig:ddrate0}
\end{figure}

The results in this section are in order of increasing interaction strength,
from lowest to highest, starting with the non-interacting regime in
Fig.~\ref{fig:ddrate0}, where Fig.~\ref{fig:ddrate0}(a) shows the rates of
coverage decrease for the overall surface (solid black line), and the
contributions from each group of sites (symbols), classified according to
their coordination number $z_i$. Figure \ref{fig:ddrate0}(b) shows a magnified
view of site contributions alone.
\begin{figure}
  \includegraphics[width=3.5in]{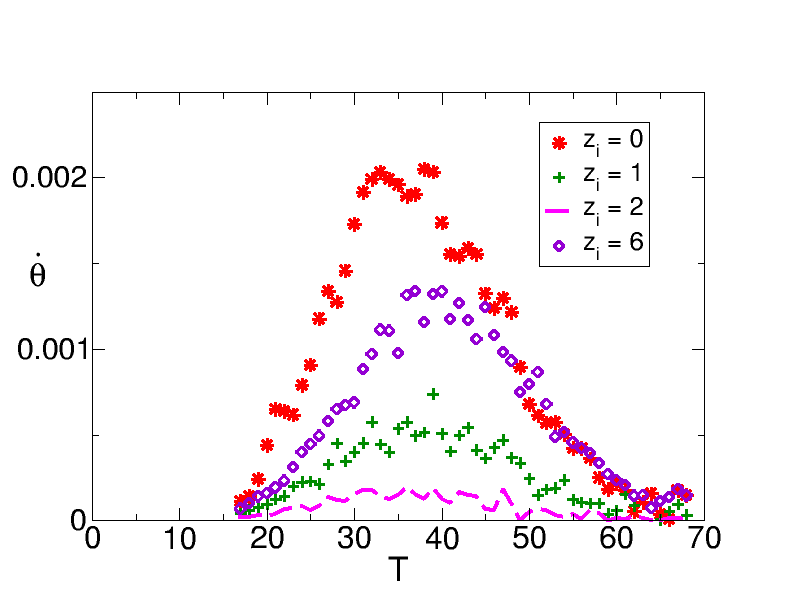}
  \caption{Desorption rates as function of temperature at $0\%$
    interaction strength for sites with $z_{i} = 0,~1,~2,~6$. Desorption from
    sites with $z_i = 0$ occurs at a faster rate than others, whereas
    desorption from sites with $z_i = 6$ occurs last. The desorption rate is
    governed mainly by a greater probability for diffusion, as the number of
    nearest neighbors $z_{i}$ increases.}
\label{fig:ddrt0}
\end{figure}

In Fig.~\ref{fig:ddrate0}(a) the peak for sites with $z_{i} = 0$ reaches a
maximum at a slightly lower temperature than the rest, as it shifts leftward
with respect to the others, this is easier to see in the zoomed-in
plot of Fig.~\ref{fig:ddrt0}. In Fig.~\ref{fig:ddrate0}(b) it can
  also be seen that there are very mild differences in the peak temperatures for
sites with $3 \leq z_{i} \leq 5$, as the $z_i = 5$ peak is shifted slightly rightward,
indicating a slightly slower desorption rate.  The $z_{i} = 6$ sites have the
  slowest desorption rate, since the corresponding peak is shifted toward high temperature
with respect to the rest, this is easier to see in Fig.~\ref{fig:ddrt0}. 
As mentioned before, the rates vary from site
to site only because of varying site coordination, which results in some sites
having `options' to diffuse to an available nearest neighbor, slowing down
desorption from those locations. The small differences between non-interacting
Arrhenius plots at low $T$ in Fig.~\ref{fig:Noint}(b) are
likely caused by the fast initial desorption from sites with $z_{i} = 0$, but
the differences are not so prominent when the binding energies are the same at
all sites.  The various rates result in varying frequencies of desorption
events, and, although very mildly for this system, this affects the frequency
of desorption events of the preexponential factor $\nu$ in a way that is
\textit{independent of molecular interactions}.

There is some irony in how the intrinsic disorder of the amorphous surface
reduces the same randomness in the desorption process that makes this regime
trivial in the crystalline lattice. A different site distribution would perhaps
accentuate differences and might cause the Polanyi-Wigner equation of desorption
to fail to model the overall desorption rate
\cite{Talbot:07,Talbot:08,Zuniga:12a}.

\begin{figure}
    \includegraphics[width=3.5in]{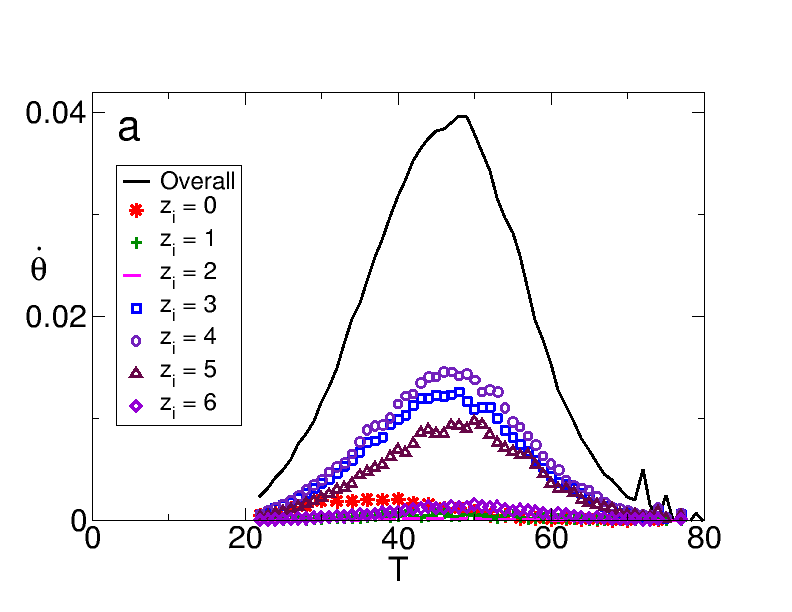}
    \includegraphics[width=3.5in]{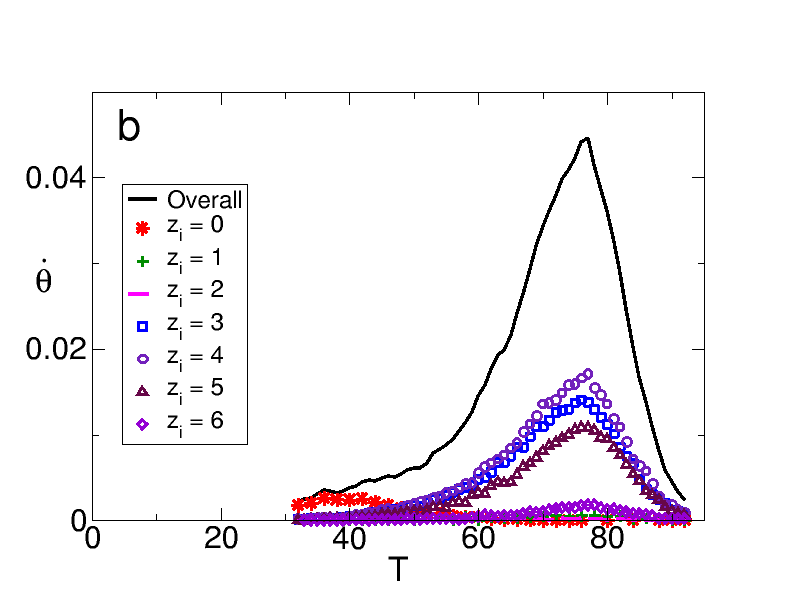}
    \includegraphics[width=3.5in]{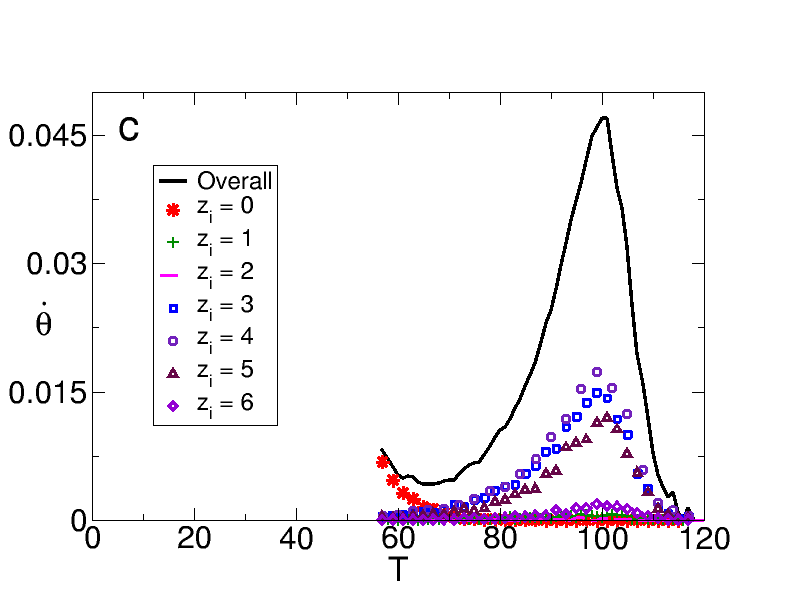}
  \caption{Desorption rates as function of temperature at (a) $10\%$
    (b) $50\%$  and (c) $90\%$ interaction strengths.
    The overall rate is the solid black line. Site contributions are  classified according 
    to their coordination number
    $z_{i}$, and plotted in symbols.}
\label{fig:ddrate105090}
\end{figure}
\begin{figure}
   \includegraphics[width=3.5in]{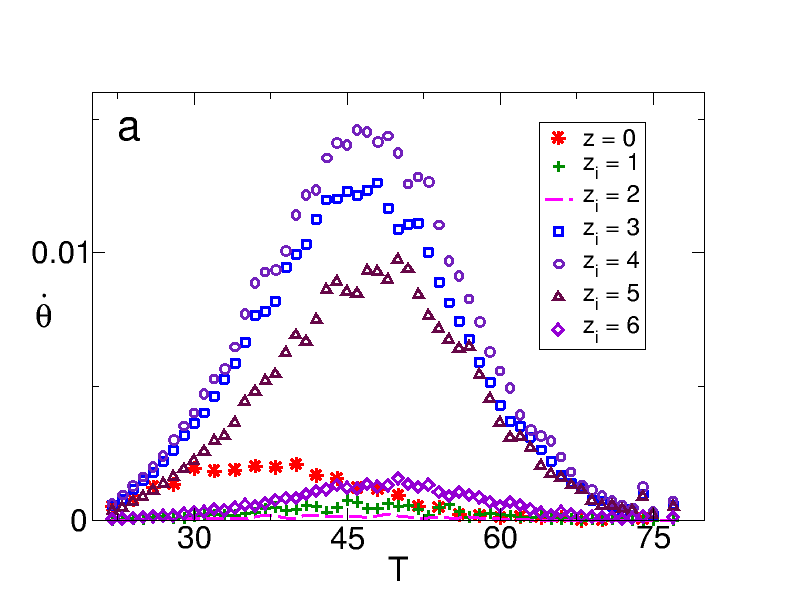}
    \includegraphics[width=3.5in]{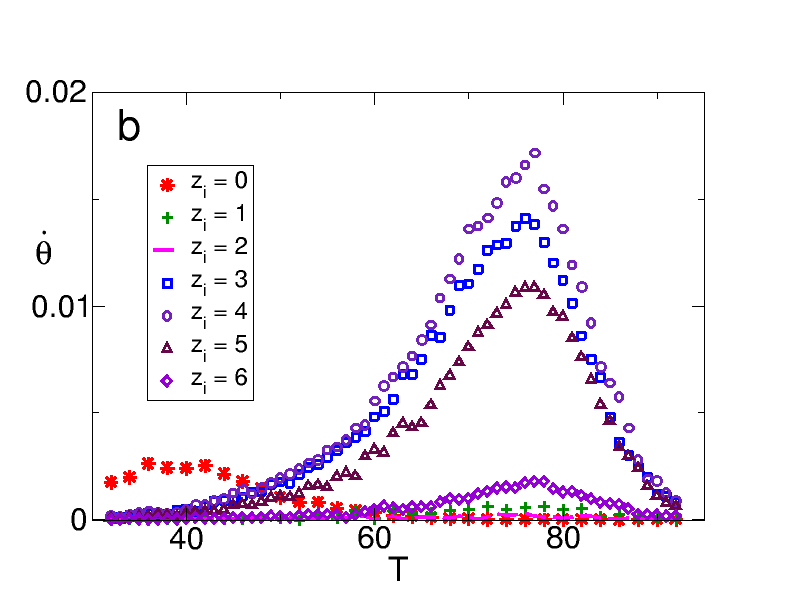}
  \includegraphics[width=3.5in]{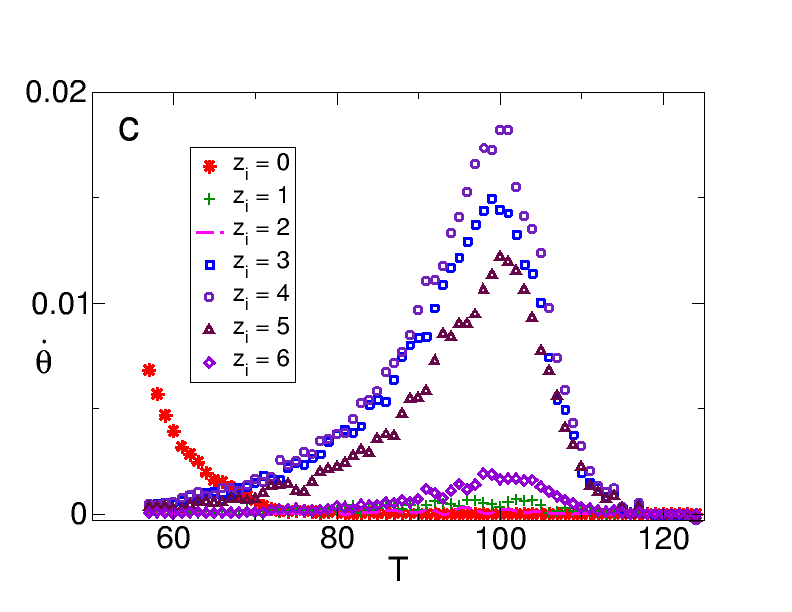}
    \caption{ Magnified view of site contributions at (a) $10\%$, (b) $50\%$ and 
      (c) $90\%$ interaction strengths. The first step in the
     process is fast desorption from sites with $z_{i} = 0$ nearest neighbors. The
      rest of the peaks shift rightward with respect to the rattler one,
      indicating slower desorption rates. Note that the temperature axis does not
     start at $0$ to provide a magnified view.}
\label{fig:rtzoom105090}
\vspace{-0.2cm}
\end{figure}

In the interacting regime adsorption sites have
the additional energetic contribution of attractive lateral interactions.
This increases the effective desorption barrier to be overcome by particles,
which also depends on the coordination number of the site at which they are located,
and also results in an energetic heterogeneity that yields various desorption rates,
as seen in Fig.~\ref{fig:ddrate105090}.
Sites with $z_i = 6$ naturally provide the largest desorption barrier.

In Figs.~\ref{fig:ddrate105090}(a), (b) and (c), the desorption rate
peak for $z_{i} = 0$ remains to the left of the rest, but changes in shape.
At $10\%$ and $50\%$ interaction strengths (Figs.~\ref{fig:ddrate105090}(a) and (b),
repsectively) the $z_i = 0$ peaks 
retain the typical thermal desorption shape. However,
for $\epsilon = 0.5 E_b$ its leftmost end does not start at $0$ on the
abscissa, which is the reason why the overall rate curve in Fig.~\ref{fig:ddrate105090}(b)
does not start at $0$ either. At $90\%$ interaction strength
(Fig.~\ref{fig:ddrate105090}(c)) the $z_i = 0$ peak 
is seen to produce the lefmost `tail' on the left end of the overall 
rate. In this interaction regime this initial step takes place
very fast, like flash desorption.

These features can be seen more clearly in
Figs.~\ref{fig:rtzoom105090}(a), (b) and (c), where the magnified picture
of site contributions in the $10\%$, $50\%$ and 
$90\%$  interaction regime, respectively, are shown.

 The differences in the shapes of the $z_i = 0$ peaks
can be attributed to the initial temperature $T_0$ of the simulation run,
which, for purposes of comparison, was selected to match that of the same
interaction strength regime of the crystal, given that the Arrhenius plots are
constructed as a function of $\frac{1}{T}$.

\begin{figure}
    \includegraphics[width=3.5in]{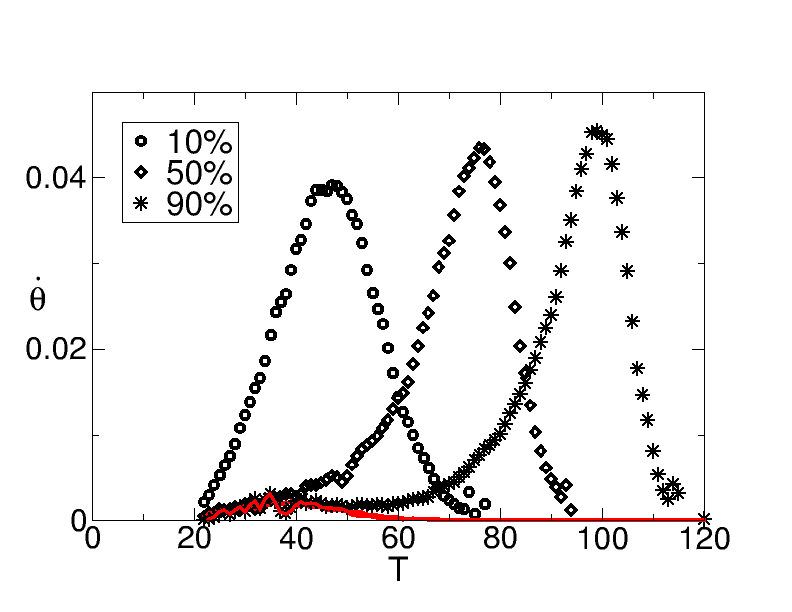}
    \caption{Overall rates of desorption for $10\%$, $50\%$ and $90\%$
      interaction strength (symbols) and rates of desorption from rattler
      sites (red lines).  The rates of desorption from rattler sites depend
      only on initial temperature. The red curves on the left overlap when a
      common value of $T_0$ is used for all desorption runs. This feature, in
      addition to faster rates from sites with $z_{i} < 4$, yields an overall
      faster rate of desorption in the amorphous surface.}
\label{fig:lowT}
\vspace{-0.2cm}
\end{figure}

If $T_0$ is set to the same low value for all regimes of interaction strength
in the amorphous surface, the $z_i = 0$ desorption rate peak spans the same
temperature range and has the same typical shape in all cases. This can be
seen in Fig.~\ref{fig:lowT}, where only the $z_i = 0$ and overall desorption
rates are shown. This demonstrates that this desorption rate depends
solely on $T_0$, and is independent of lateral interactions.

The transient variations in the Arrhenius parameters throughout
the desorption process will be explored next.

\subsection{Activation Energy}

As part of a systematic study of the kinetic compensation effect in thermal
desorption, this section presents the numerically calculated transient
variations in the energy of activation per site, per iteration, as a function of coverage
$\theta$. Site contributions to the overall $E_a$ curves 
were also calculated. As previously mentioned, this study is intended to explore how a
change in an `experimental' parameter (in this case the surface configuration)
may result in a KCE, an IKR, or both.

\begin{figure}
    \includegraphics[width=3.5in]{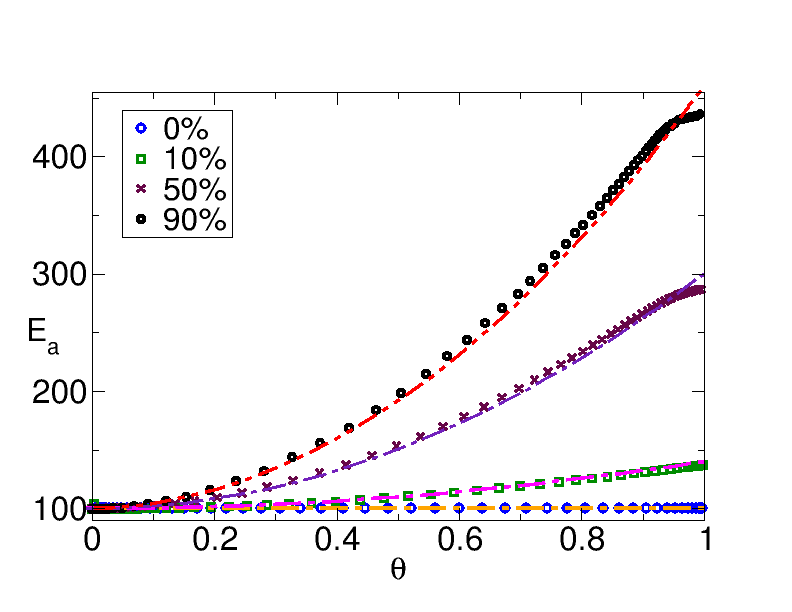}

    \caption{Effective activation energy $E_{a}$ as a function of coverage
      $\theta$ at $0\%$, $10\%$, $50\%$ and $90\%$ interaction
      strength. Comparison between the crystalline and disordered
      surfaces.  $E_a$ for the disordered surface exhibtis the same functional
      form as that of the crystal, except for a small initial difference in the presence
      of interactions, more visible at $50\%$ and $90\%$ interaction
      strengths.}
\label{fig:Ea}
\end{figure}

\begin{figure}
    \includegraphics[width=3.5in]{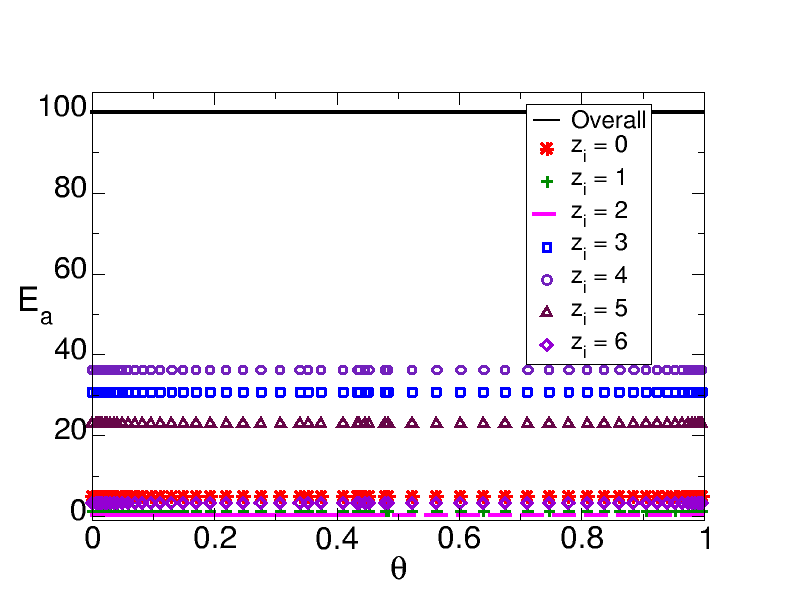}
    \caption{Activation energy $E_{a}$, per site as function of coverage
      $\theta$, for the amorphous surface, at $0\%$ lateral interaction
      strength. In this regime all site contributions ti $E_a$ remain constant.}
\label{fig:Ea0}
\end{figure}

\begin{figure}
      \includegraphics[width=3.5in]{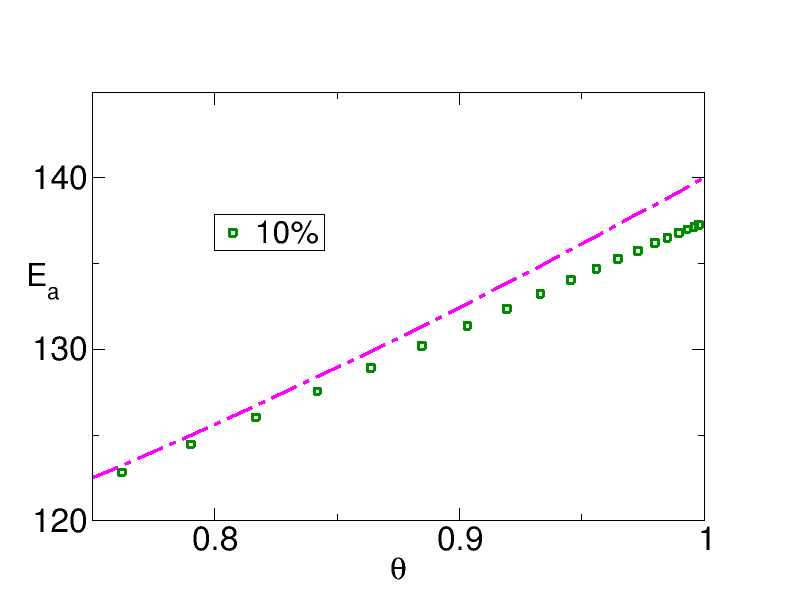}
\caption{Magnified view of the initial stages of the variation of $E_{a}$
      as function of coverage $\theta$, at $10\%$ interaction strength.
      The initial difference between $E_a$ curves is also observed
     in this interaction regime.}
\label{fig:Ea10zoom}
\end{figure}

\begin{figure}
    \includegraphics[width=3.5in]{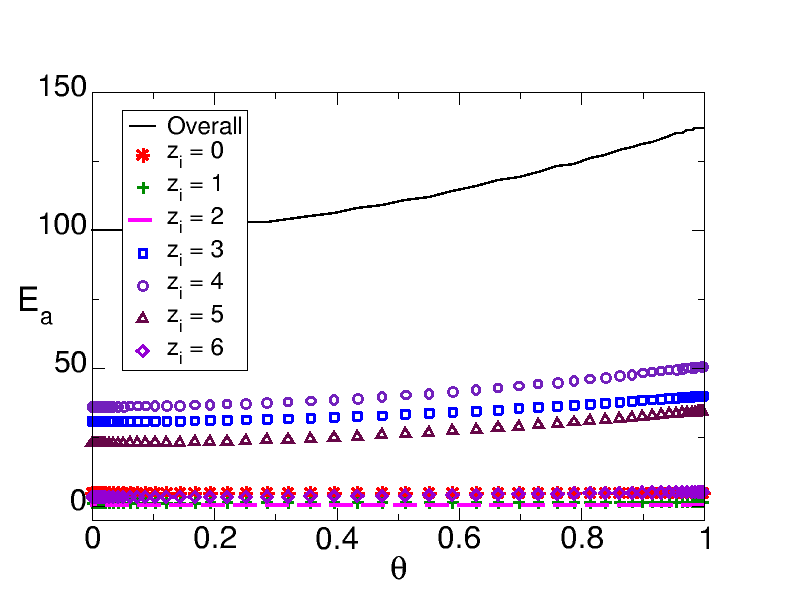}
\caption{Activation energy per site as function of coverage in the amorphous surface
         at $10\%$ interaction strength. In the presence of lateral interactions
       only the contribution from sites with $z_i = 0$ (red stars) remains constant, while site
      contributions for $z_i > 0$ (symbols) vary in a similar fashion as the overall curve (black 
     solid line).}
\label{fig:Ea10}
\end{figure}

\begin{figure}
    \includegraphics[width=3.5in]{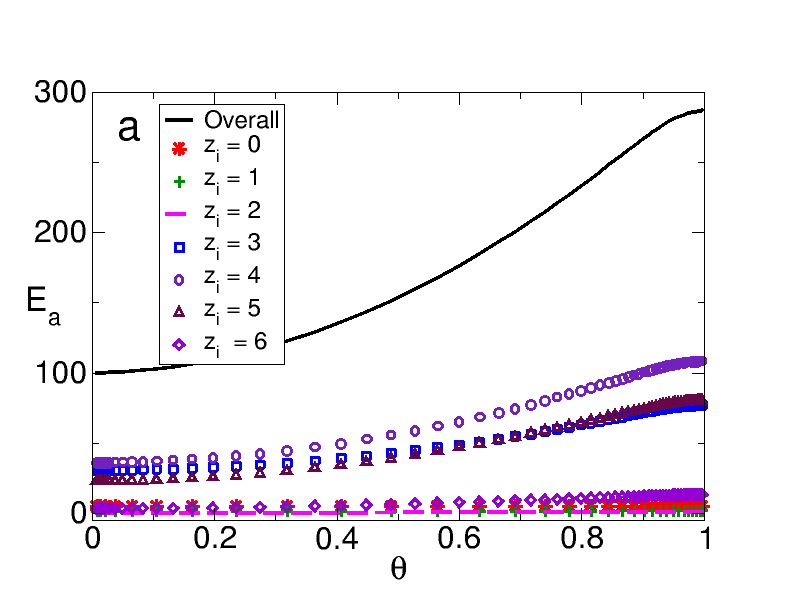}
    \includegraphics[width=3.5in]{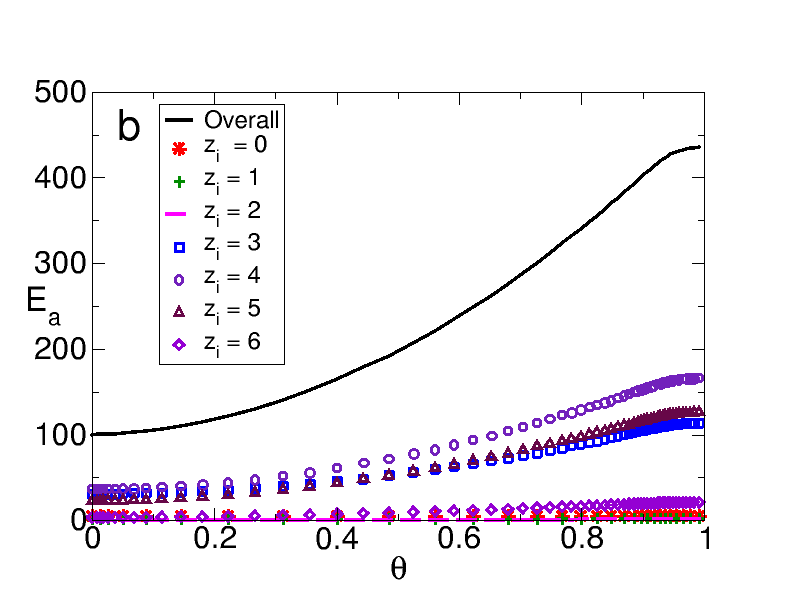}
    \caption{Activation energy per site as function of $\theta$ in the
      amorphous surface at (a) $50\%$ and (b) $90\%$ interaction strengths. 
      Site contributions (symbols) show that $E_a$ decreases from its initial magnitude
    more slowly as $z_i$ increases. The initial `stagnation' is produced by slower
    decrease from the initial $E_a$ magnitude for sites with $3 \le z_i \le 5$.
    The rattler contribution remains constant in all regimes of 
    interaction strength.}
\label{fig:Ea5090}
\end{figure}

Figure \ref{fig:Ea}
shows a comparison between the numerically calculated $E_a$ curves throughout the desorption process
from the crystalline (lines) and the amorphous surface (symbols), for all regimes
of interaction strength studied here.
At $0\%$ interaction strength, $E_a$ remains constant through the 
entire process, as expected, and all site contributions, shown in Fig.~\ref{fig:Ea0},
remain constant as well.
Fig.~\ref{fig:Ea} also shows that in the interacting regime 
the behavior of $E_a$ from the disordered surface did not deviate much from
the behavior observed for the crystal in \cite{Zuniga:18}, except for an initial numerical difference, 
followed by a brief `stagnation' at the initial stage of the desorption process (at low $T$ and high coverage),
where $E_a$ in the amorphous surface briefly decreases from its largest magnitude at a slower pace,
before regressing to the behavior observed in 
the crystal. These initial differences are more visible at
$50\%$ and $90\%$ interaction strengths, but also
occur at $10\%$ interaction strength, and can be seen in
the zoomed-in picture in Fig.~\ref{fig:Ea10zoom}.
The initial numerical difference arises because the effective average
coordination number $z_i$ per site at monolayer coverage is $3.87$ for the amorphous surface, 
and $4$ for the crystal. The initial `stagnation' occurs because 
rattler desorption dominates this initial stage, with a net energetic contribution to the 
overall $E_a$ curve of $0$. And, while this step takes place, 
desorption from sites with $z_i = 3$ to $z_i = 6$ occurs more
slowly, which is why the overall $E_a$ curve also decreases more slowly.
This is not very evident in Fig.~\ref{fig:Ea10}, but can be seen in 
Figs.~\ref{fig:Ea5090}(a) and (b),
where site contributions to the overall $E_a$ curve at every recorded stage of the process
are plotted.
In Figs.~\ref{fig:Ea5090}(a) and (b), the $z_i= 3$ to $z_i = 5$ curves have
the most visible effect on the shape of the overall $E_a$ curve.
The $z_i = 1$ and $z_i = 2$ curves are not mentioned, this is
because there is a very small number of these sites, and their net 
contribution to the overall is rather negligible.
For $z_i = 6$ the corresponding $E_a$ curve changes even more sluggishly than the rest, but
a smaller number of these sites compared to those for $z_i = 3$ to $z_i = 5$ 
causes their net effect to average out.

\subsection{Preexponential factor}
\begin{figure}
    \includegraphics[width=3.5in]{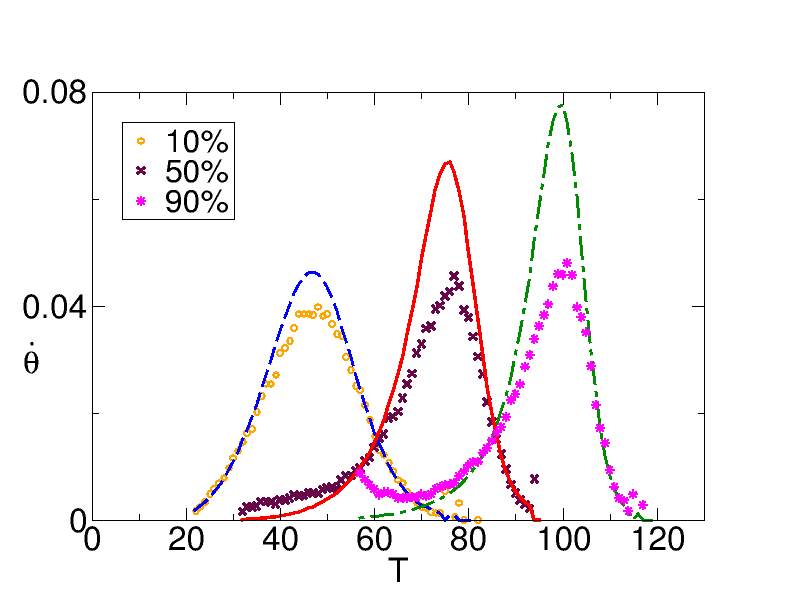}
    \caption{Comparison of rates of desorption in the amorphous surface. The
      solid lines represent the peaks calculated with the Polanyi-Wigner
      equation, Eq.~\ref{eq:PW}, using the numerical data for $E_a$, $\theta$,
      and $T$, and setting $\nu = 1$. The symbols represent the time
      derivative taken directly from the coverage data. The difference between
      the peaks indicates a compensation effect due to variations in the
      prefactor.}
\label{fig:peaks}
\vspace{-0.2cm}
\end{figure}
\begin{figure}
    \includegraphics[width=3.5in]{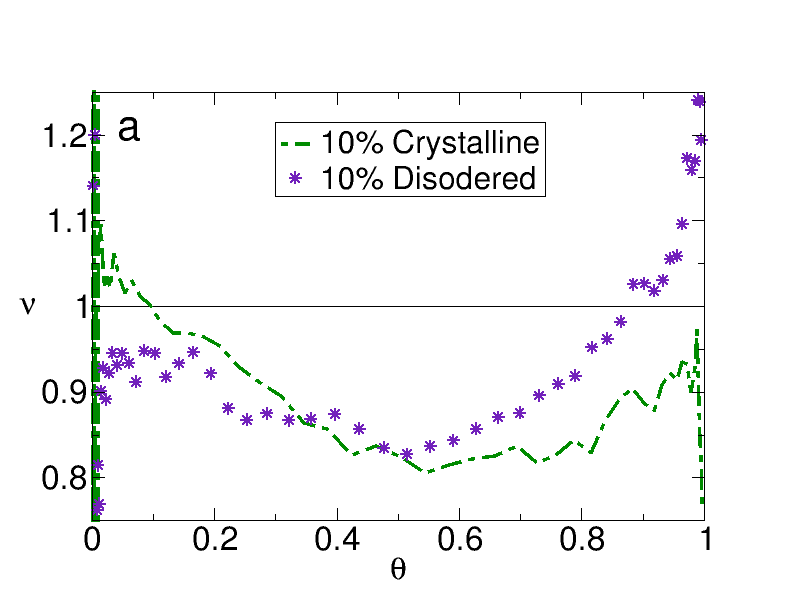}
    \includegraphics[width=3.5in]{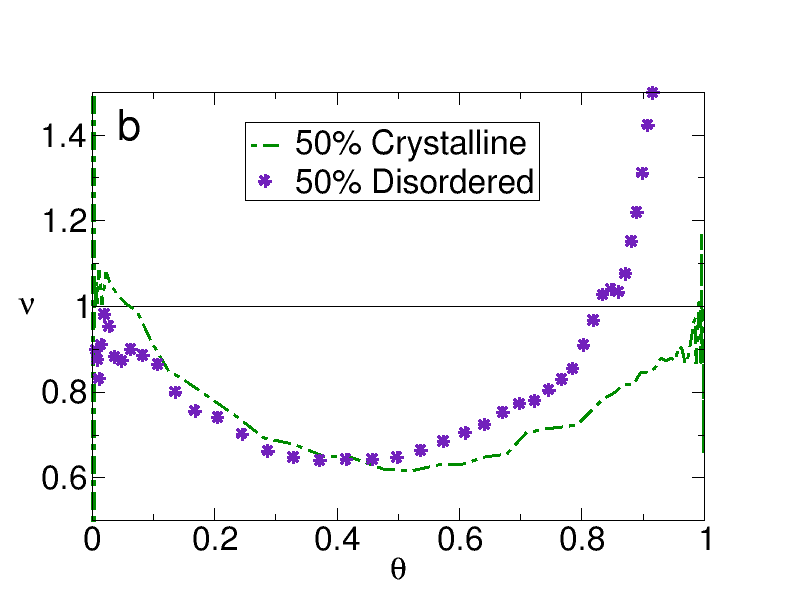}
    \includegraphics[width=3.5in]{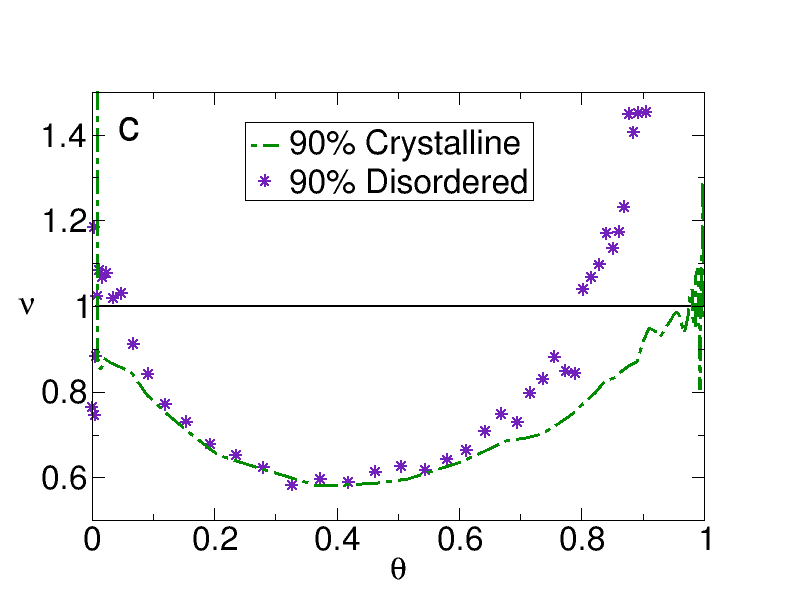}
    \caption{Preexponential factor as a function of coverage for $10\%$,
      $50\%$, and $90\%$ interaction strength in increasing order from top to
      bottom. The prefactor $\nu$ from the amorphous configuration exhibits
      very large variations, above its standard value of $1$, at high coverage
      values.  At about $80\%$ fractional surface coverage, $\nu$ for the
      amorphous surface drops below $1$ and rescinds to the behavior observed
      for the crystal, where the decrease is likely originated by a decrease
      in the frequency of desorption events and configurational entropy.}
\label{fig:nuvscov}
\vspace{-0.2cm}
\end{figure}
The preexponential factor in the amorphous surface is calculated
 by dividing two desorption rate peaks.  One is obtained directly from the time
derivative $\dot{\theta}$ of the surface coverage decrease data, the result is
represented by symbols in Fig.~\ref{fig:peaks}, and the analytically
calculated rate, using Eq.~\ref{eq:PW} for order $1$, with the numerical data
for $E_a$, $\theta$ and $T$, and setting $\nu = 1$, represented by straight
lines in Fig.~\ref{fig:peaks},  in the same manner as was done for the crystal in \cite{Zuniga:18}.
The differences between the rates indicate that there is some
amount of compensation, which is greater at the peak temperature. Also, the
prefactor contribution is what generates the leftmost `tail' at $90\%$
interaction strength, and it is also the reason why the overall peak does not start
at $0$ on the abscissa at $50\%$ interaction energy.

The calculated prefactors for each interaction strength regime are plotted in
Fig.~\ref{fig:nuvscov}, and overlayed with the results from the crystal from
\cite{Zuniga:18} for comparison.  In Figs.~\ref{fig:nuvscov}(a), (b) and (c) it is seen that,
in the interacting regime, $\nu$ from the amorphous configuration exhibits
large variations at the beginning of the desorption process, and then drops
below $1$ at around $80\%$ fractional surface coverage.  At this point, the
behavior of $\nu$ closely resembles that of the prefactor in the crystal.  The
large variations at high coverage values show the fast initial desorption from
rattler sites, this mostly affects the frequency component of $\nu$, as it is
a result of the increase in the frequency of desorption events. When $\nu$
drops from unity, there is a slowdown in the number of those same events, due
to island formation \cite{Gunther:14} and increased effective desorption
barriers.  This clustering also causes a decrease in configurational entropy,
but in the amorphous surface there is the additional factor of sites with
$z_{i} = 0$ being unavailable for reoccupation after particles desorb, and
also that the different $z_{i}$ values from site to site make some locations
easier to reoccupy than others. The latter effect does not seem too
pronounced in this configuration, since the variations in $\nu$ are very close to 
those in the crystal. Perhaps this is due to all sites having the same binding energy.
It may become more prominent for a
different site distribution, or if the lattice is energetically heterogeneous
\cite{Zuniga:12a}.

\subsection{Kinetic compensation effect for the amorphous surface}

\begin{figure}
    \includegraphics[width=3.5in]{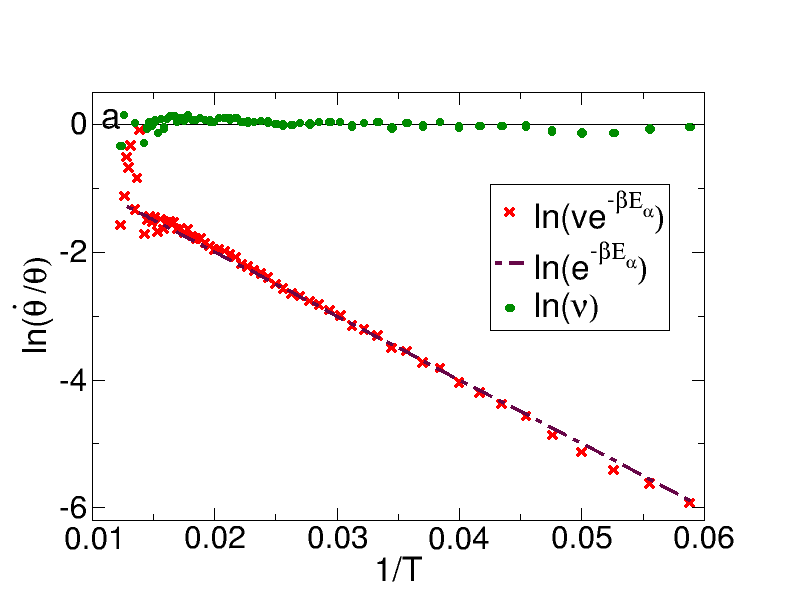}
    \includegraphics[width=3.5in]{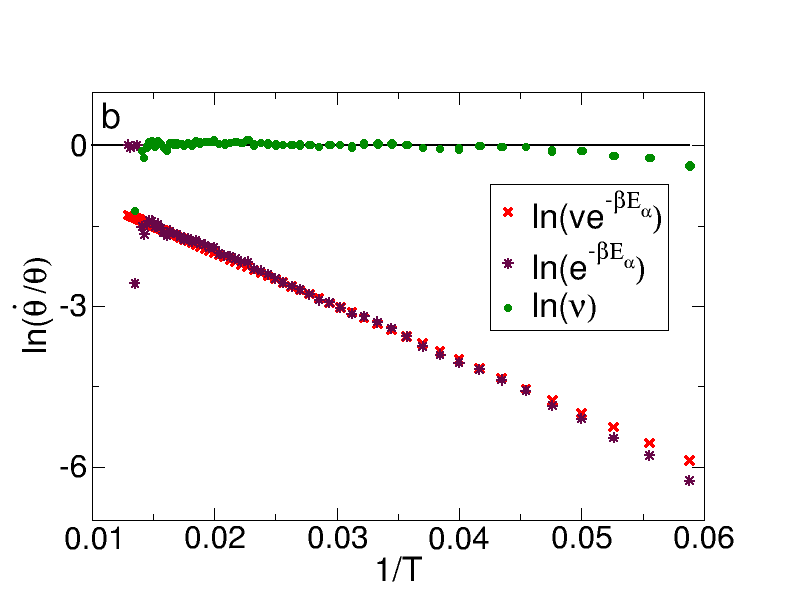}
    \caption{Arrhenius plot (red stars) and separate contributions from $E_a$
      (purple dashed line) and $\nu$ (green dots) at $0\%$ interaction
      strength for (a) the amorphous surface, and (b) the crystal.  
     The $E_a$ contribution almost overlaps completely with
     the Arrhenius plot in this regime, except for a portion at the
      beginning of the desorption process, which happens when the prefactor
      slightly drops below the $0$ axis in the figure.}
\label{fig:aplotsz40}
\vspace{-0.2cm}
\end{figure}

Here the contributions of $E_a$ and $\nu$ to the overall Arrhenius plots for
the disordered surface are quantified. This is done by directly comparing the two
sides of the natural logarithm of the Arrhenius equation, which gives the following:
\beq 
\ln{k} = \frac{-E_a}{T} + \ln{\nu}
\label{eq:LnArrh}
\eeq
All terms in Eq.~\ref{eq:LnArrh} in are calculated
with the numerical data in the previous sections, and plotted as
function of $\frac{1}{T}$. Here $k = -\frac{\dot{\theta}}{\theta}$.  The first
set of results correspond to the non-interacting regime
(Fig.~\ref{fig:aplotsz40}).
In the absence of lateral interactions, the $E_a$ contribution almost completely overlaps
with the overall Arrhenius plot, but there are some small differences at the
beginning of the process, where there appears to be a slight drop of the
$\ln{\nu}$ plot from $0$.  This can be seen with the reference axis added in
Fig.~\ref{fig:aplotsz40}(a).  This may be purely numerical, therefore the same
results for the crystal are plotted in Fig.~\ref{fig:aplotsz40}(b) for
comparison.  A drop is observed here too, however it is more pronounced.  The
drop could also be caused by a brief initial slowdown in the frequency of
desorption attempts due to low initial temperature. 
\begin{figure}
    \includegraphics[width=3.5in]{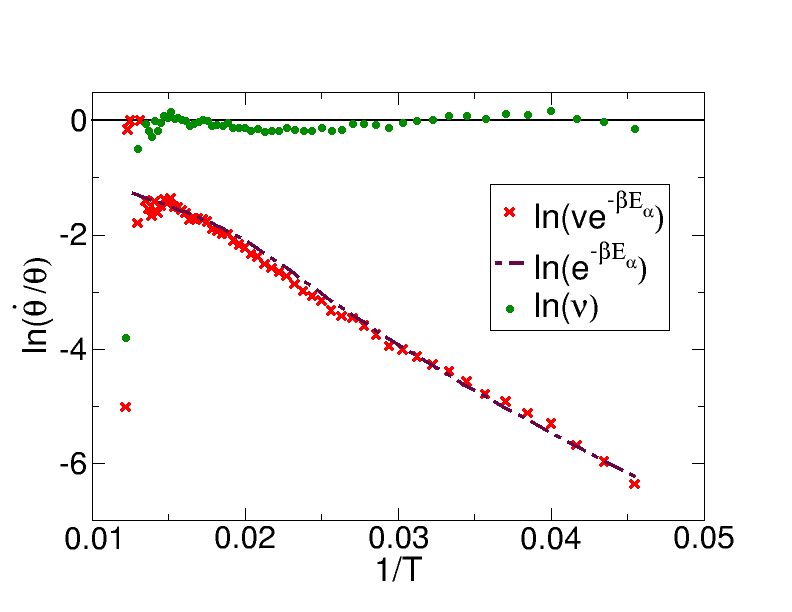}
\caption{Arrhenius plot (red stars), separate contributions from 
$E_a$ (purple dashed line), and $\nu$ (green dots) at $10\%$ interaction strength.
The $E_a$ contribution almost completely overlaps with the Arrhenius plot in this regime, but there
are some differences, showing
a small level of \textit{partial} compensation due to $\ln{\nu}$. The KCE and IKR are usually attributed 
to weak molecular interactions, where the Arrhenius plots can be fitted to a straight line.}
\label{fig:aplotsz410}
\vspace{-0.2cm}
\end{figure}

At $10\%$ interaction strength, in Fig.~\ref{fig:aplotsz410}, the Arrhenius
plot is very mildly curved, but can still be fit to a straight line.  The
prefactor contribution $\ln{\nu}$ shows small variations that are consistent
with its transient behavior observed in
Fig.~\ref{fig:nuvscov}(a): first there is a variation in the opposite direction of
$E_a$, where $\nu$ acquires very large values, and after this a small drop
from zero, similar to that observed in the crystal.  These features are not so visible
here, but the trend is easier to see as the interaction strength parameter
$\epsilon$ is increased.

\begin{figure}
    \includegraphics[width=3.5in]{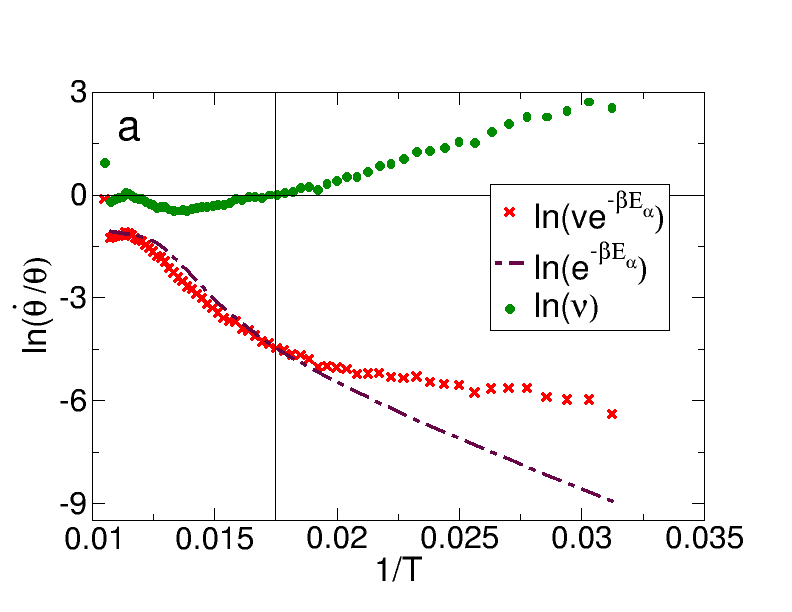}
    \includegraphics[width=3.5in]{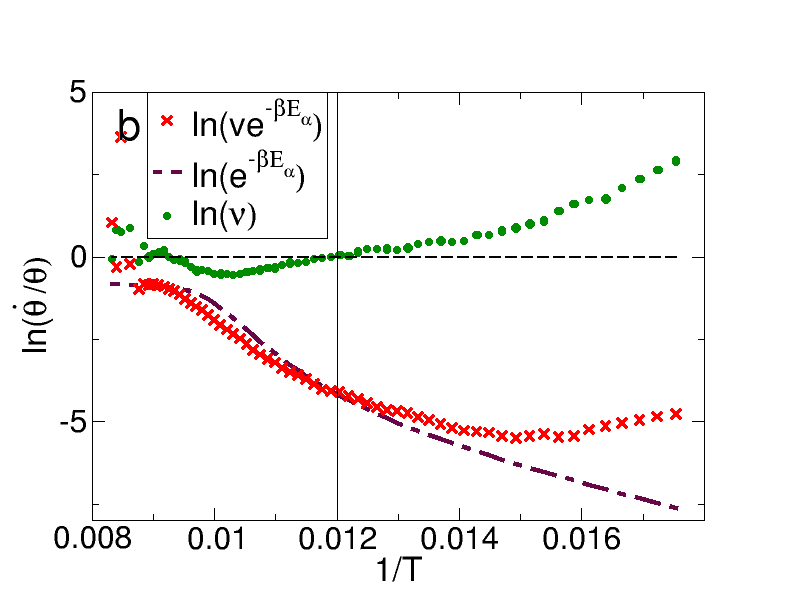}
   \caption{Arrhenius plot (red stars), separate contributions from $E_a$
      (purple dashed line), and $\nu$ (green dots) at (a) $50\%$ 
      and (b) $90\%$ interaction
      strength.  There is a large difference between the $E_a$ contribution
      and the overall Arrhenius plot, because of the large initial variation
     in the preexponential factor. This $\ln{\nu}$ contribution varies in
      the opposite direction of $E_a$, and mitigates the effect of the enhanced 
      effective desorption barrier due to interactions, but is 
      \textit{independent} of $E_a$. After this step $\nu$
       weakly compensates for changes in $E_a$.}
\label{fig:aplotsz45090}
\end{figure}
In the $50\%$ and $90\%$ regimes
(Figs.~\ref{fig:aplotsz45090}(a) and (b), respectively), it can be seen that the large initial
variation in the prefactor (Figs.~\ref{fig:nuvscov}(b) and (c)) generates the
increased curvature of the Arrhenius plot, while the $E_a$ contribution by
itself results in an Arrhenius plot that is slightly curved, a signature of
the variable energy of activation \cite{Vyazokin:16}, and closely resembles
the $E_a$ contributions and Arrhenius plots observed for the crystal
\cite{Zuniga:18}.

Whenever there is a decrease in $E_a$ (i.e., more negative and therefore corresponds to a stronger
binding) that results in a decrease in configurational entropy, 
it is said that there is a compensation effect \cite{LiuGuo:01,Piguet:11}. This implies that the
parameters move in the same direction and offset each other.  If the
parameters move in the opposite direction, for example if $E_a$ becomes
less negative for repulsive interactions, the configurational entropy would
still decrease, in this case mainly because of site exclusion. Then this is referred to as the
less explored \textit{anti compensation effect}, where in principle the parameters would
vary in the opposite direction \cite{Piguet:11}. The fast initial
desorption resembles the effect of a net repulsive interaction, even though
repulsive interactions are not included explicitly in this study.
Nevertheless, it mitigates the slowdown in frequency of desorption events due
to the enhanced desorption barriers, so it is somehow compensating for these
changes \textit{in the opposite direction}. However, the fast initial
desorption is \textit{independent of lateral interactions} and,  as seen in Fig.~\ref{fig:lowT} 
it depends only on
the initial temperature, and raises the question of whether this can be
characterized as any type of compensatory behavior at all, given the lack mutual
dependence.  Nevertheless, this factor plays a role in determining the overall rate, and would
be omitted if the functional characterization of $\nu$ is done based on 
the \textit{a priori} assumption of complete compensation between $E_a$ and $\ln{\nu}$
to satisfy Eq.~\ref{eq:IKR}. A similar point is made in ref. \cite{Chodera:13}.

Many authors question the validity of a strong linear correlation on the basis of the percentage
error with which the parameters can be obtained
from experimental data \cite{Perez:16,Barrie:12a,Cornish:02,Chodera:13}. However,
given that the breakdown of the contributions in this section
show that, at least in this system, no complete compensation occurs, 
 it also seems reasonable to ask, if the slope of Eq.~\ref{eq:IKR}
does not match the crossing temperature, what is the information that a strong linear correlation 
between the parameters provides?
Perhaps it is a semi-empirical relation between apparent Arrhenius parameters
 which could be useful to characterize the effects of experimental
changes in activated processes, much how apparent Arrhenius parameters allow for the semi-empirical
characterization of rates \cite{Barrie:11,Agrawal:86}. 


\section{Conclusions}

In summary, attemping to characterize $\nu$ purely as a function of $E_a$, and vice versa, in order to
force the parameters to mutually compensate each other and fit the linear correlation in Eq.~\ref{eq:IKR}
may omit important factors that determine the overall rate of a process.
For the amorphous configuration in this study, the most prominent factor is the fast desorption from sites with $0$ 
nearest neighbors. This initial step is \textit{independent of lateral interactions} and yields variations
in $\nu$ in the opposite direction to those in $E_a$.
This raises the question of 
whether this can be characterized as a compensation effect,
yet the increase in the frequency of desorption events 
mitigates the effect of a  larger potential barrier to be overcome by adsorbates in the 
presence of attractive interactions. 
After the initial desorption step, $\nu$ is observed to
partially compensate for changes in $E_a$, in 
a similar manner to that observed in \cite{Zuniga:18}.
This type of weak compensation is also favored by other authors \cite{LiuGuo:01,Chodera:13}.

The KCE is defined as the linear correlation between apparent Arrhenius parameters
in Eq.~\ref{eq:IKR}, but it is only possible to extract constant values for $E_a$ and $\ln{\nu}$ from the 
slope and $y-$intercept, respectively, of a fairly linear Arrhenius plot. Even if that is the case for weak
interactions, the results here, and those in \cite{Zuniga:18}, show that for this system the 
level of compensation is actually weak, and the parameters do not offset one another,
even at the point of isokinetic equilibrium.  It therefore seems reasonable to
ask what information can be obtained, other than the compensation temperature, 
from a linear correlation between $E_a$ and $\ln{\nu}$. It could be a semi-empirical way to characterize the 
effects of changing a particular parameter and, as mentioned by 
other authors \cite{Perez:16,Chodera:13}, the size of the errors in measurement should
be considered.
Compensation was also observed when the interaction
strength increases, which means that a curved Arrhenius plot does not necessarily
exclude this effect.


\end{document}